\documentclass[12pt]{article}
\usepackage[cp1251]{inputenc}
\usepackage{epsf}

\title{Distribution of valence quarks in hadrons in QCD: theoretical
method of calculations.}
\author{B.L.Ioffe\\
 Institute of Theoretical and Experimental Physics,\\
B.Cheremushkinskaya 25, 117218 Moscow,Russia}
\date{}
\begin{document}

\maketitle

\newcommand{\be}{\begin{equation}}
\newcommand{\ee}{\end{equation}}

\def\la{\mathrel{\mathpalette\fun <}}
\def\ga{\mathrel{\mathpalette\fun >}}
\def\fun#1#2{\lower3.6pt\vbox{\baselineskip0pt\lineskip.9pt
\ialign{$\mathsurround=0pt#1\hfil##\hfil$\crcr#2\crcr\sim\crcr}}}


\begin{abstract}
The general method for calculation of valence quark distributions
in hadrons at intermediate $x$ is presented. The imaginary part of
virtual photon forward scattering amplitude on  quark current with
hadron quantum number  is considered in the case, when initial and
final virtualities   of the current $p^2_1$ and $p^2_2$ are
different, negative and large: $\mid p^2_1\mid, \mid p^2_2\mid\gg
R^{-2}_c$, where $R_c$ is confinement radius. The operator product
expansion (OPE) in $p^2_1,p^2_2$ up to dimension 6 operators is
performed. Double dispersion representations in $p^2_1, p^2_2$ of
the amplitude in terms of physical states contributions are used.
Equalling them to those calculated in QCD by OPE the desired sum
rules for quark distributions in mesons are found. The double
Borel transformations are applied to the sum rules, killing
non-diagonal transition terms, which deteriorated the accuracy in
the previous calculations of quark distributions in nucleon.
Leading order perturbative corrections are accounted. Valence
quark distributions in pion, longitudinally and transversally
polarized $\rho$-mesons and proton are calculated at intermediate
$x$, $0.2 \la x \la 0.7$ and normalization points $Q^2 = 2-5
~GeV^2$ with no fitting parameters. In cases of pion and proton
the results are in agreement with found correspondingly from the
data on the Drell-Yan process and deep inelastic scattering.
Valence quark distributions in transversally  and longitudinally
polarized $\rho$-mesons are essentially different one from another
and also differ from those in pion.

\end{abstract}

\vspace{5mm}

{\bf \large 1. ~Introduction}

\vspace{3mm}
 Quark and gluon distributions in hadrons are not fully
understood in QCD. QCD predicts the evolution of these
distributions with $Q^2$ in accord with the
Dokshitzer-Gribov-Lipatov-Altarelli-Parisi (DGLAP) [1]-[3]
equations, but not the initial values from which this evolution
starts. The standard way of determination of quark and gluon
distributions in nucleon is the following [4-8] (for the recent
review see [9]). At some $Q^2 = Q^2_0$ (usually, at low or
intermediate $Q^2 \sim 2-5 GeV^2$) the form of quark (valence and
sea) and gluon distributions is assumed and characterized by the
number of free parameters. Then, by using DGLAP equations, quark
and gluon distributions are calculated at all $Q^2$ and $x$ and
compared with the whole set of the data on deep inelastic
lepton-nucleon scattering (sometimes also with prompt photon
production, jets at high $p_{\bot}$ etc; for pion -- with the data
on Drell-Yan process). The best fit for the parameters is found
and, therefore, quark and gluon distributions are determined at
all $Q^2$, including their initial values $q(Q^2_0, x)$,
~$g(Q^2_0, x)$. Evidently, such an approach is not completely
satisfactory from theoretical point of view - it would be
desirable to determine the initial distribution directly from QCD.

I present you based on QCD method of the calculation of $u$ and
$d$ valence quark distributions in hadrons at low $Q^2\sim 2-5$
GeV$^2$ and intermediate $x$. The idea of the method was suggested
in [10] and developed in [11-13]. Recently, the method had been
improved and valence quark distributions in pion [14] and
transversally and longitudinally polarized $\rho$-meson [15] had
been calculated, what was impossible in the initial version of the
method. For valence quark distributions in proton the improved
method [16] gave a much better agreement with experiment, than the
earlier version [11]. The idea of the approach (in the improved
version) is to consider the imaginary part (in $s$-channel) of a
four-point correlator $\Pi(p_1, p_2, q, q^{\prime})$ corresponding
to the forward scattering of two quark currents, one of which has
the quantum numbers of hadron of interest  and the other is
electromagnetic. It is supposed that virtualities of the photon
$q^2, q^{\prime 2}$ and hadron currents $p^2_1, p^2_2$ are large
and negative $\vert q^2 \vert = \vert q^{\prime 2} \vert \gg \vert
p^2_1 \vert,~ \vert p^2_2 \vert \gg R^{-2}_c$, where $R_c$ is the
confinement radius. It was shown in [11] that in this case the
imaginary part in $s$-channel $[s = (p_1 + q)^2]$ of $\Pi(p_1,
p_2; q_1, q^{\prime})$ is dominated by a small distance
contribution at intermediate $x$. (The standard notation is used:
$x$ is the Bjorken scaling variable, $x = -q^2/2 \nu$~, $\nu = p_1
q)$.  The proof of this statement, given in [11], is based on the
fact that for the imaginary part of the forward scattering
amplitude the position of the closest to zero singularity in
momentum transfer is determined by the boundary of the Mandelstam
spectral function and is given by the equation
\be
t_0 = -4 \frac{x}{1-x} p^2 \label{1} \ee Therefore, if $\vert p^2
\vert$ is large and $x$ is not small, then even at $t = 0$ (the
forward amplitude) the virtualities of intermediate states in
$t$-channel are large enough for OPE to be applicable. So, in the
mentioned above domain of $q^2, q^{\prime 2}$,~ $p^2_1, p^2_2$ and
intermediate $x$~ $Im \Pi(p,p_2; q, q^{\prime})$ can be calculated
using the perturbation theory and the operator product expansion
in both sets of variables $q^2 = q^{\prime 2}$ and $p^2_1, p^2_2$.
The approach is inapplicable at small $x$ and $x$ close to 1. This
can be understood for physical reasons. In deep inelastic
scattering at large $\vert q^2 \vert$ the main interaction region
in space-time is the light-cone domain and longitudinal distances
along the light-cone are proportional to $1/x$ and become large at
small $x$ [17,18]. For OPE validity it is necessary for these
longitudinal distances along light-cone to be also small, that is
not the case at small $x$. At $1 - x \ll 1$ another condition of
applicability of the method is violated. The total energy square
$s = Q^2(1/x - 1) + p^2_1$~ $Q^2 = -q^2$ is not large at $1 - x
\ll 1$. Numerically, the typical values to be used below are  $Q^2
\sim 5 GeV^2$, ~ $p^2_1 \sim - 1 GeV^2$. Then, even at $x \approx
0.7$, $s \approx 1 GeV^2$, i.e., at such $x$ we are in the
resonance region. So, one may expect beforehand, that our method
could work only up to $x \approx 0.7$. The inapplicability of the
method at small and large $x$ manifests itself in the blow-up of
higher order terms of OPE. More precise limits on the
applicability domain in $x$ will be found from the magnitude of
these terms.

The further procedure is common for QCD sum rules. On one hand the
four-point correlator $\Pi(p_1, p_2; q, q^{\prime})$ is calculated
by perturbation theory and OPE.On the other hand, the double
dispersion representation in $p^2_1, p^2_2$ in terms of physical
states contributions is written for the same correlator and the
contribution of the lowest state is extracted using the Borel
transformaion. By equaling these two expression the desired quark
distribution is found.

\bigskip

{\bf \large 2. ~The method}

\vspace{3mm} Consider the forward 4-point correlator:
\be
 \Pi (p_1, p_2; q, q^{\prime}) = -i \int~ d^4x d^4y d^4z e^{ip_1x
+ iqy - ip_2z} \times \langle 0 \vert T \left \{j^h(x),~
j^{el}(y),~ j^{el}(0),~ j^h(z) \right \} \vert 0 \rangle \label{4}
\ee Here $p_1$ and $p_2$ are the initial and final momenta carried
by hadronic current $j^h$, $q$ and $q^{\prime} = q + p_1 - p_2$
are the initial and final momenta carried by virtual photons.
(Lorenz indeces are omitted). It will be very essential for us to
consider non-equal $p_1, p_2$ and treat $p^2_1, p^2_2$ as two
independent variables. However, we may put $q^2 = q^{\prime 2} =
q^2$ and $t = (p_1 - p_2)^2 = 0$. We are interested in imaginary
part of $\Pi(p^2_1, p^2_2, q^2, s)$ in $s$ channel:
\be
Im \Pi (p^2_1, p^2_2, q^2, s) = \frac{1}{2i} \Biggl [\Pi(p^2_1,
p^2_2, q^2, s + i \varepsilon) - \Pi(p^2_1, p^2_2, q^2, s -
i\varepsilon) \Biggr ] \label{5} \ee In order to construct
representation of $Im \Pi(p^2_1, p^2_2, q^2, s)$ in terms of
contributions of physical states, let us write for $Im \Pi(p^2_1,
p^2_2, q^2, s)$ the double dispersion relation in $p^2_1, p^2_2$:
$$ Im \Pi(p^2_1, p^2_2, q^2, s) = a(q^2, s) + \int
\limits^{\infty}_0 ~\frac{\varphi(q^2, s, u)}{u - p^2_1} du + \int
\limits^{\infty} _{0} \frac {\varphi (q^2, s, u)}{u - p^2_2} du$$
\be
+ \int \limits ^{\infty}_{0} du_1~ \int \limits ^{\infty} _{0}
du_2~ \frac {\rho(q^2, s, u_1, u_2)}{(u_1 - p^2_1) (u_2 - p^2_2)}
\label{6} \ee The second and the third terms in the right-hand
side (rhs) of (\ref{6}) may
 be considered as subtraction terms to the last one -- the properly double
 spectral representation. The first term in the rhs of (\ref{6}) is the
 subtraction term to the second and third ones. Therefore, (\ref{6}) has the
 general form of the double spectral representation with one subtraction in
 both variables -- $p^2_1$ and $p^2_2$.  Apply the double Borel
 transformation in $p^2_1, p^2_2$ to (\ref{6}). This transformation kills
 three first terms in rhs of (\ref{6}) and we have
\be
{\cal{B}}_{M^2_1} {\cal{B}}_{M^2_2}~ Im \Pi(p^2_1, p^2_2, q^2, s)
= \int \limits ^{\infty} _{0} du_1~ \int \limits ^{\infty}_{0}
du_2 \rho (q^2, s, u_1, u_2) exp \Biggl [-\frac{u_1}{M^2_1} -
\frac{u_2}{M^2_2} \Biggr ] \label{7} \ee The integration region
over $u_1, u_2$ may be divided into 4 areas:

I. $u_1 < s_0;~~ u_2 < s_0$

II.$u_1 < s_0; ~~ u_2 > s_0$

III. $u_1 > s_0; ~~ u_2 < s_0$

IV. $u_1, u_2 > s_0$

Using the standard QCD sum rule model of hadronic spectrum and the
hypothesis of quark-hadron duality, i.e. the model with one lowest
resonance plus continuum, one may see, that area I corresponds to
resonance contribution. Spectral density in this area can be
written as
\be
\rho(u_1, u_2, x, Q^2) = g^2_h \cdot 2\pi F_2 (x, Q^2) \delta(u_1
- m^2_h) \delta(u_2 - m^2_h), \label{8} \ee where $g_h$ is defined
as
\be
\langle 0 \vert j_h \vert h \rangle = g_h \label{9} \ee (For
simplicity we consider the case of the Lorenz scalar hadronic
current.) If in $Im \Pi(p_1, p_2, q, q^{\prime})$ the structure,
proportional to $P_{\mu} P_{\nu}$ $[P_{\mu} = (p_1 +
p_2)_{\mu}/2]$ is considered, then in the lowest twist
approximation $F_2(x, Q^2)$ is the structure function, depending
on the Bjorken scaling variable $x$ and weakly on $Q^2 = -q^2$.

In area (IV), where both variables $u_{1,2}$ are far from
resonance region, the non-perturbative effects may be neglected,
and as usual in sum rules, the spectral function of hadron state
is described by the bare loop spectral function $\rho^0$ in the
same region
\be
\rho(u_1, u_2, x) = \rho^0(u_1, u_2, x) \label{10} \ee In areas
(II),(III) one of the variables is far from the resonance region,
but other is in the resonance region, and the spectral function in
this region is some unknown function $\rho = \psi(u_1, u_2, x)$,
which corresponds to transitions like $h \to continuum$. The areas
II-III contributions  are exponentially suppressed, and, using the
standard hypothesis of quark-hadron duality, we may estimate them
as a bare loop contribution in the same integration region.
 Equating
physical and QCD representation of $\Pi$ and taking into account
cancellation of appropriate parts in the left and right sides, one
can write the following sum rules:
$$ Im~ \Pi^0_{QCD} + \mbox{Power~ correction} = 2 \pi F_2 (x, Q^2)
g^2_h e ^{-m^2_h(\frac{1}{M^2_1} + \frac{1}{M^2_2})} $$
\be
Im \Pi^0_{QCD} = \int \limits^{s_0}_{0}\int
\limits^{s_0}_{0}\rho^0 (u_1, u_2, x) e^{- \frac{u_1 + u_2}{2M^2}}
\label{12} \ee It can be shown (see below), that for bare loop
diagram $\psi(u_1, u_2, x)) \sim \delta(u_1 - u_2)$, and, as a
consequence, the areas II and III contributions are zero in our
model of hadronic spectrum.

It is worth mentioning that if we would consider the forward
scattering amplitude from the beginning, put $p_1 = p_2 = p$ and
perform Borel transformation in $p^2$, then unlike (5), the
contributions of the second and third terms in (\ref{6}) would not
 vanish. They just correspond to the non-diagonal
transition matrix elements  and are proportional to
\be
\langle 0 \vert j^h \vert h^{\ast} \rangle ~ \frac{1}{p^2 -
m^{*2}_h}~ \langle h^{\ast} \vert j^{el}(x) j^{el} (0) \vert h
\rangle ~ \frac{1}{p^2 - m^2_h} \langle h \vert j^h \vert 0
\rangle \label{13} \ee From decomposition
\be
\frac{1}{p^2 - m^{*2}_h} ~ \frac{1}{p^2 - m^2_h} =
\frac{1}{m^{*2}_h - m^2_h} \Biggl (\frac{1}{p^2 - m^{*2}_h} -
\frac{1}{p^2 - m^2_h} \Biggr ) \label{14} \ee it is clear that in
this case (\ref{13}) may contribute to the second (or third) term
in (\ref{6}) and after Borel transformation the contribution of
the second term in (\ref{14}) has the same Borel exponent
$e^{-m^2_h/M^2}$ as the lowest resonance contribution. The only
difference is in pre-exponent factors: they are $1/M^2$ in front
of the resonance term and Const. in front of non-diagonal terms.
This difference was used in order to get rid of non-diagonal
terms:  application of the differential operator
$(\partial/\partial (1/M^2)e^{m^2_h/M^2}$ to the sum rule kills
the Borel non-suppressed nondiagonal terms, but deteriorates the
accuracy and shrinks the applicability domain of the sum rule
(particularly, the domain in $x$, where the sum rule is valid).

\newpage

 {\bf \large 3.~ Quark distributions in pion}

\vspace{3mm} It is enough to find the distribution of valence
$u$-quark in $\pi^+$, since $\bar{d}(x) = u(x)$. The most suitable
hadronic current in this case is the axial current
\be
j_{\mu 5} = \bar{u} \gamma_{\mu} \gamma_5 d \label{19} \ee In
order to find the $u$-quark distribution, the electromagnetic
current is chosen as $u$-quark current with the unit charge
\be
j^{el}_{\mu} = \bar{u} \gamma_{\mu} u \label{20} \ee The bare loop
Fig.1 contribution is given by
$$ Im~ \Pi_{\mu \nu \lambda \sigma} = -\frac{3}{(2 \pi)^2}~
\frac{1}{2}~ \int~ \frac{d^4 k}{k^2} ~ \frac{1}{(k + p_2 - p_1)^2}
\delta [(q + k)^2
 ] \delta [p_1 - k)^2 ]
$$
\be
\times Tr \left \{\gamma_{\lambda} \hat{k} \gamma_{\mu}(\hat{k} +
\hat{q}) \gamma_{\nu} (\hat{k} + \hat{p}_2 - \hat{p}_1)
\gamma_{\sigma} (\hat{k} - \hat{p}_1) \right \} \label{21} \ee It
is convenient to introduce
\be P = (p_1 + p_2)/2,~~~  r = p_1 - p_2, ~ r^2 = 0 \label{17} \ee
The tensor structure, chosen to construct the sum rule is a
structure proportional to $P_{\mu} P_{\nu} P_{\lambda}
P_{\sigma}/\nu$.  The reasons are the following. The results of
the QCD sum rules calculations are more reliable, if invariant
amplitude at kinematical structure with maximal dimension is used.
Different $p_1 \not=p_2$ are important  only in denominators,
where they allow one to separate the terms in dispersion
relations. In numerators one may restrict oneself to consideration
of terms proportional to 4-vector $P_{\mu}$ and ignore the terms
$\sim r_{\mu}$. Then the factor $P_{\mu} P_{\nu}$ provides the
contribution of $F_2(x)$ structure function and the factor
$P_{\lambda} P_{\sigma}$ corresponds to contribution of spin zero
states. (The factor $1/\nu$ is scaling  factor : $w_2 = F_2/\nu$.)

Let us use the notation
\be
\Pi_{\mu \nu \lambda \sigma} = (P_{\mu} P_{\nu} P_{\lambda}
P_{\sigma}/\nu) \Pi (p^2_1, p^2_2, x) + ... \label{22} \ee Then
$Im \Pi (p^2_1, p^2_2, x)$ can be calculated from (\ref{21}) and
the result is [14]:
\be
Im \Pi (p^2_1, p^2_2, x) = \frac{3}{\pi} x^2 (1 - x) \int \limits
^{\infty} _{0}~ du_1 \int \limits ^{\infty} _{0} du_2 ~
\frac{\delta(u_1 - u_2)}{(u_1 - p^2_1)(u_2 - p^2_2)} \label{23}
\ee The matrix element of the axial current between vacuum and
pion state is well known
\be
\langle 0 \vert j_{\mu 5} \vert \pi \rangle = i p_{\mu} f_{\pi}
\label{24} \ee where $f_{\pi}= 131 MeV$. The use of (\ref{12}),
(\ref{23}), (\ref{24}) gives the sum rule for valence $u$-quark
distribution in pion in the bare loop approximation [14]:
\be
u_{\pi}(x) = \frac{3}{2 \pi^2}~\frac{M^2}{f^2_{\pi}} x (1-x) (1 -
e^{-s_0/M^2}) e^{m^2_{\pi}/M^2}, \label{25} \ee where $s_0$ is the
continuum threshold. In ref.[14] the following corrections to
(\ref{25}) were accounted:

1. Leading order (LO) perturbative corrections, proportional to
$ln(Q^2/\mu^2)$ , where $\mu^2$ is the normalization point. In
what follows the normalization point will be chosen to be equal to
the Borel parameter $\mu^2 = M^2$.

2.  Higher order terms of OPE. Among the latter, the dimension-4
correction, proportional to gluon condensate $\langle 0 \vert
\frac{\alpha_s}{\pi} G^n_{\mu \nu}~ G^n_{\mu\nu} \vert 0 \rangle$
was first accounted, but it was found that its contribution to the
sum rule vanishes after double borelization. There are two types
of vacuum expectation values (v.e.v) of dimension 6: one, where
only gluonic fields enter:
\be
\frac{g_s}{\pi} \alpha_s f^{abc} \langle 0 \vert G^{a}_{\mu \nu}~
G^b_{\nu \lambda}~ G^c_{\lambda \mu} \vert 0 \rangle \label{26}
\ee and the other, proportional to four-quark operators
\be
\langle 0 \vert \bar{\psi} \Gamma \psi \cdot \bar{\psi} \Gamma
\psi \vert 0 \rangle \label{27} \ee It was shown in [14] that
terms of the first type cancel in the sum rule and only terms of
the second type survive. For the latter one may use the
factorization hypothesis which reduces all the terms of this type
to the square of quark condensate.

 A remark is in order here. As was mentioneed in the Introduction, the
 present approach is invalid at small and large $x$. No-loop 4-quark
 condensate contributions are proportional to $\delta(1-x)$ and
  cannot be
 accounted. In the same way, the diagrams, which can be considered as a
 radiative corrections to those, proportional to $\delta(1-x)$, must be also
 omitted.

 All dimension-6 power corrections to the sum rule were calculated in
 ref.'s 14,15 and the final result is given by:
$$ xu_{\pi}(x) =
\frac{3}{2\pi^2}\frac{M^2e^{m^2_{\pi}/M^2}}{f^2_{\pi}}x^2(1-x)
\Biggl [ \Biggl ( 1+ \Biggl (\frac{a_s(M^2)\cdot
ln(Q^2_0/M^2)}{3\pi}\Biggr )$$ $$\times \Biggl ( \frac{1+4x
ln(1-x)}{x}- \frac{2(1-2x)ln x}{1-x}\Biggl ) \Biggl )\cdot
(1-e^{-s_0/M^2}) $$
\be
-\frac{4\pi \alpha_s(M^2)\cdot 4\pi \alpha_s a^2}{(2\pi)^4 \cdot
3^7\cdot 2^6\cdot M^6} \cdot \frac{\omega(x)}{x^3(1-x)^3}\Biggr ],
\label{28} \ee where $\omega(x)$ is 4-order polynomial in $x$,
(for its explicit form see [14]),
\be a = -(2 \pi)^2 \langle 0 \vert  \bar{\psi} \psi \vert 0
\rangle \label{29} \ee $u_{\pi}(x)$ may be used as an initial
condition at $Q^2 = Q^2_0$ for solution of QCD evolution equations
(DGLAP equations).

In numerical calculations we choose: the effective LO QCD
parameter $\Lambda^{LO}_{QCD} = 250 MeV,~ Q^2_0 = 2 GeV^2, ~
\alpha_s a^2 (1 GeV^2) = 0.34 GeV^6$. The value of the latter is
taken from the recent best fit of hadronic $\tau$-decay data [19]
and corresponds to  $\alpha_s\langle
\overline{\psi}\psi\rangle^2=(2.2\pm 0.7)\cdot 10^{-4}$ GeV$^6$.
The continuum threshold was varied in the interval $0.8 < s_0 <
1.2 GeV^2$ and it was found, that the results only slightly depend
on it. The analysis of the sum rule (\ref{28}) shows, that it is
fulfilled in the region $0.15 < x < 0.7$; the power corrections
are less than 30\% and the continuum contribution is small ($<
25$\%). The stability in the Borel mass parameter $M^2$ dependence
in the region $0.4 GeV^2 < M^2 < 0.6 GeV^2$ is good. The result of
our calculation of valence distribution in pion $x
u_{\pi}(x,Q^2_0)$ is shown  in Fig.2.

Suppose, that at small $x \la 0.15 ~~ u_{\pi}(x) \sim 1/\sqrt{x}$
according to Regge behaviour and at large $x \ga 0.7~~ u_{\pi}(x)
\sim (1-x)^2$ according to quark counting rules. Then, matching
these functions with (\ref{28}), one may find the numerical values
of the first and second moments of $u$-quark distribution
\be
{\cal{M}}_1 = \int \limits^1 _0 u_{\pi} (x) dx \approx 1.1
\label{31} \ee
\be
{\cal{M}}_2 = \int \limits^1_0 xu_{\pi} (x) dx \approx 0.24
\label{32} \ee ${\cal{M}}_1$ has the meaning of the number of
$u$-quarks in $\pi^+$ and should be ${\cal{M}} = 1$. The deviation
of (\ref{31}) from 1 characterizes the accuracy of our
calculation. ${\cal{M}}_2$ has the meaning of the part of pion
momentum carried by valence $u$-quark. Therefore, valence $u$ and
$\bar{d}$ quarks are carrying about 40\% of the total momentum. In
Fig.2 are  plotted also the valence $u$-quark distribution found
in [4b] by fitting the data on production of $\mu^+\mu^-$ and
$e^+e^-$ pairs in pion-nucleon collisions (Drell-Yan process).

\bigskip

{\bf \large 4.~ Quark distributions in $\rho$-meson}

\vspace{3mm} The choice of hadronic current is evident

\be
j_{\mu}^{\rho} = \overline{u}\gamma_{\mu}d \label{33} \ee The
matrix element $\langle \rho^+\mid j^{\rho}_{\mu}\mid 0 \rangle$
is given by

\be
\langle \rho^+\mid j^{\rho}_{\mu}\mid 0 \rangle =
\frac{m^2_{\rho}}{g_{\rho}}e_{\mu} \ee where $m_{\rho}$ is the
$\rho$-meson mass, $g_{\rho}$ is the $\rho-\gamma$ coupling
constant, $g^2_{\rho}/4\pi=1.27,~e_{\mu}$  is the $\rho$-meson
polarization vector. Consider separately $u$-quark distributions
in longitudinally and transversally polarized $\rho$-mesons. It
was shown in [15], that for determination of $u$-quark
distribution in longitudinal $\rho$-meson the most suitable tensor
structure is that, proportional to $P_{\mu} P_{\nu}
P_{\sigma}P_{\lambda}$, while $u$-quark distribution in transverse
$\rho$  can be found by considering the invariant function at the
structure $-P_{\mu}P_{\nu}\delta_{\lambda\sigma}$. In case of
longitudinal $\rho$-meson the tensor structure, which is separated
is the same, as in the case of pion. Since at $m_q=0$ bare loop
contributions for vector and axial hadronic currents are equal,
the only difference from the pion case is in the normalization. It
can be shown, that $u$-quark distribution in longitudinal
$\rho$-meson can be found from (22) by substitutions $m_{\pi}\to
m_{\rho}$, $f_{\pi}\to m_{\rho}/g_{\rho}$ and therefore
\be
xu^L_{\rho}(x) = \frac{f^2_{\pi}}{m^2_{\rho}}
g^2_{\rho}e^{(m^2_{\rho}-m^2_{\pi})/M^2} xu_{\pi}(x),\ee where
$xu_{\pi}(x)$ is given by (22). (The numerical values of $M^2$ and
$s_0$ differ in the cases of pion and longitudinal $\rho$-meson).
After separation of the invariant function at the structure
$P_{\mu}P_{\nu}\delta_{\lambda\sigma}$ in box diagram we find
$u$-quark distribution in transversally polarized $\rho$-meson in
bare loop approximation:
\be
u^T_{\rho}(x)_0 = 3 \Biggl [ \frac{1}{2} - x (1-x)\Biggr ]\equiv
\frac{3}{2}\varphi_0(x) \ee $u^T_{\rho}(x)_0$ has a minimum at
$x=\frac{1}{2}$, unlike $u_{\pi}(x)_0$ and $u^L_{\rho}(x)_0$.
Another essential difference in comparison with pion and
longitudinal $\rho$ arises due to the fact that gluon condensate
and $\langle G^3\rangle$ operator contributions are nonvanishing
in the sum rule for $u^T_{\rho}(x)$.

Valence $u$-quark distribution in transversally polarized
$\rho$-meson is given by
$$ xu^T_{\rho}(x) = \frac{3}{8 \pi^2} g^2_{\rho}
e^{m^2_{\rho}/M^2}~ \frac{M^4}{m^4_{\rho}} x \left \{\varphi_0(x)
E_1 \Biggl (\frac{s_0}{M^2} \Biggr ) \Biggl [ 1 + \frac{1}{3 \pi}
ln \Biggl (\frac{Q^2_0}{M^2} \Biggr ) \alpha_s (M^2) \Biggl (
\frac{(4x-1)}{\varphi_0(x)} + \right.  $$ $$ + 4 ln (1-x) -
\frac{2(1 - 2x + 4x^2) ln x}{\varphi_0(x)} \Biggr ) \Biggr ]-
\frac{\pi^2}{6}~ \frac{\langle 0 \vert(\alpha_s/\pi) G^2 \vert 0
\rangle} {M^4 x^2}  $$ $$ +\frac{1}{2^8 \cdot 3^5 M^6 x^3 (1-x)^3}
\langle 0 \vert g^3 f^{abc} G^a_{\mu \nu} G^b_{\nu \lambda}
G^c_{\lambda \mu} \vert 0 \rangle \xi (x) $$
\be
\left. +\frac{\alpha_s(M^2) (\alpha_s a^2)}{2^5 \cdot 3^8 \pi^2
M^6 x^3 (1 - x)^3 }\chi (x) \right\} \ee where $\xi(x)$ and
$\chi(x)$ are polynomials in $x$ [15]. In numerical calculations
the values  of parameters were chosen: $M^2=1$ GeV$^2$ (this value
is in the middle of stability interval in $M^2$), $s_0=1.5$
GeV$^2$, $Q^2_0=4$ GeV$^2$, $\langle 0 \mid (\alpha_s/\pi)G^2\mid
0 \rangle=0.006$ GeV$^2$ [20], $\langle G^3\rangle$ vacuum
expectation value was estimated according to instanton model [21].
The effective instanton radius, was taken to be equal $\rho_c=0.5$
fm. The other parameters are the same as in calculation of
$xu_{\pi}(x)$. On Fig.3 are plotted $u^L_{\rho}(x),u^T_{\rho}(x)$
and,  for comparison, $u_{\pi}(x)$. As is seen from the Figure,
the shape of $u^T_{\rho}(x)$ is quite different from $u_{\pi}(x)$
and $u^L_{\rho}(x)$ and indicates, that  perhaps, $u^T_{\rho}(x)$
has two humps. $u^L_{\rho}(x)$ also has an interesting feature:
the second moment of $u$-quark distribution is equal to 0.4. It
means, that the part of $\rho^L$ momenta, carried by valence --
$u$ and $\overline{d}$ quarks comprises 80\% and for gluons and
sea quarks is left about 20\% only  -- a very unusual situation !

The main physical conclusion is: the quark distributions in pion
and $\rho$-meson have not to much in common. The specific
properties of pion, as a Goldstone boson manifest themselves  in
different quark distributions in comparison with $\rho$.

\bigskip

{\bf \large 5. ~Quark distributions in proton}

\vspace{3mm}
 Consider the 4-current correlator which corresponds
to the virtual photon scattering on the quark current with quantum
number of proton:
\be
T^{\mu \nu} (p_1, p_2, q, q^{\prime}) = - i \int d^4 x d^4 y d^4 z
\cdot e^{i(p_1 x+ q y - p_2 z)} \cdot \langle 0 \vert T \{ \eta
(x),~ j^{u, d}_{\mu} (y), j^{u, d}_{\nu} (0), ~ \bar{\eta} (z) \}
\vert 0 \rangle, \ee where $\eta(x)$ is the three-quark current
$\eta(x)=\varepsilon^{abc}(u^aC\gamma_{\lambda}u^b)\gamma_5
\gamma_{\lambda}d^c$ [22], $j^u_{\mu} = \bar{u} \gamma_{\mu}u$,~
$j^d_{\mu} = \bar{d} \gamma_{\mu} d$. As was shown in [11] and
generalized to nonequal $p_1, p_2$ in [16] the sum rules should be
written for invariant amplitude at the structure
$\hat{p}p_{\mu}p_{\nu}$ (it what follows it will be denoted by $Im
T/\nu$). The bare loop diagram is shown on Fig.4. The result of
its calculation after the double Borel transformation are the same
as in the case for equal $p_1 = p_2$ [11]:
\be
Im T^0_{u(d)} = \varphi^{u(d)}_0 (x) \frac{M^2}{32 \pi^3}~ E_2
(s_0/M^2) \ee where
\be
\varphi^u_0 (x) = x(1-x)^2 (1+8 x), ~~ \varphi^d_0(x) = x(1-x)^2
(1-2x), \ee $s_0$~~ \mbox{is ~ the~ continuum~ threshold}
\be
E_2(z) = 1 - (1+z+z^2/2)e^{-z} \ee The sum rules for $u$ and
$d$-quark distributions $q(x)^{u(d)}$ in bare loop approximation
are:
\be
x q(x)_0^{u(d)} = \frac{2 M^6
e^{m^2/M^2}}{\bar{\lambda}^2_N}~\varphi_0^{u(d)}(x) \cdot E_2
(\frac{s_0}{M^2}) \ee Here $\lambda_N$ is the coupling constant of
proton with the current $\langle 0 \vert \eta \vert p, r \rangle =
\lambda_N v^r(p)$, $\overline{\lambda}^2_N = 32 \pi^4\lambda^2_N$.
The first and second moments of quark distributions, with account
of proton mass sum rule in the same approximation, coincide with
quark model results, as it should be. QCD evolution was accounted
as leading  order (LO) correction to bare loop formulae (see
[16]). In OPE the contributions of $d=4$ operator (gluon
condensate) and $d=6$ operator ($\alpha_s$ times quark condensate
square) was calculated. I present here results as the ratio to
bare loop contributions:
\be
\frac{u(x)_{\langle G^2 \rangle}}{u_0(x)} = \frac{\langle
(\alpha_s/\pi) G^2 \rangle}{M^4} \cdot
\frac{\pi^2}{12}~\frac{(11+4x-31x^2)}{x(1-x)^2(1+8x)} \cdot
(1-e^{-s_0/M^2})/E_2(\frac{s_0}{M^2}) \ee
\be
\frac{d(x)_{\langle G^2 \rangle}}{d_0(x)} = - \frac{\langle
(\alpha_s/\pi) G^2 \rangle}{M^4}
\frac{\pi^2}{6}~\frac{(1-2x^2)}{x^2(1-x)^2(1+2x)}
(1-e^{-s_0/M^2})/E_2(\frac{s_0}{M^2}) \ee
\be
\frac{u(x)_{\alpha_s \langle \bar{\psi} \psi \rangle^2}}{u_0(x)} =
\frac{\alpha_s a^2 (215-867x+172x^2+288(1-x) ln 2)}{M^6 \cdot
81\pi \cdot 8x(1-x)^3(1+8x)} \frac{1}{E_2(s_0/M^2)} \ee
\be
\frac{d(x)_{\alpha_s \langle \bar{\psi} \psi \rangle^2}}{d_0(x)} =
- \frac{\alpha_s a^2(19-43x+36x^2)}{M^6 81 \pi x(1-x)^3
(1+8x)}~\frac{1}{E_2(s_0/M^2)} \ee On Fig. 5 are plotted the
obtained $u$ and $d$-quark distributions at $Q^2_0=5$ GeV$^2$ in
comparison with those, found in [4a] from the data on deep
inelastic lepton-nucleon scattering. The same values of QCD
parameters (condensates) were taken as in cases of quark
distributions in $\pi$ and $\rho$. The Borel parameter $M^2$ was
chosen in the middle of stability interval, $M^2=1.1$ GeV$^2$.

The theoretical analysis of the obtained valence quark
distributions at $Q^2 = 5 GeV^2$ showed, that $u$-quark
distribution is reliable at $0.15 < x  < 0.65$, its accuracy is
about 25\% in the middle of this interval and decreases to 50\% at
the ends of interval; $d$-quark distribution is reliable at $0.25
< x < 0.55$ with an accuracy of about 30\% in the middle and is
given by a factor of 2 at the ends of interval. In the limit of
these accuracies the theoretically calculated valence quark
distributions are in agreement with those found from deep
inelastic scattering and other hard processes data.

The research described in this publication was made possible in
part by Award No.RP2-2247 of the US Civilian and Development
Foundation for the Independent States of Former Soviet Union
(CRDF), by INTAS Grant 2000, Project 587 and by the Russian Found
of Basic Research, Grant No. 00-02-17808.

\newpage

\twocolumn

\begin{figure}
\epsfxsize=4cm \epsfbox{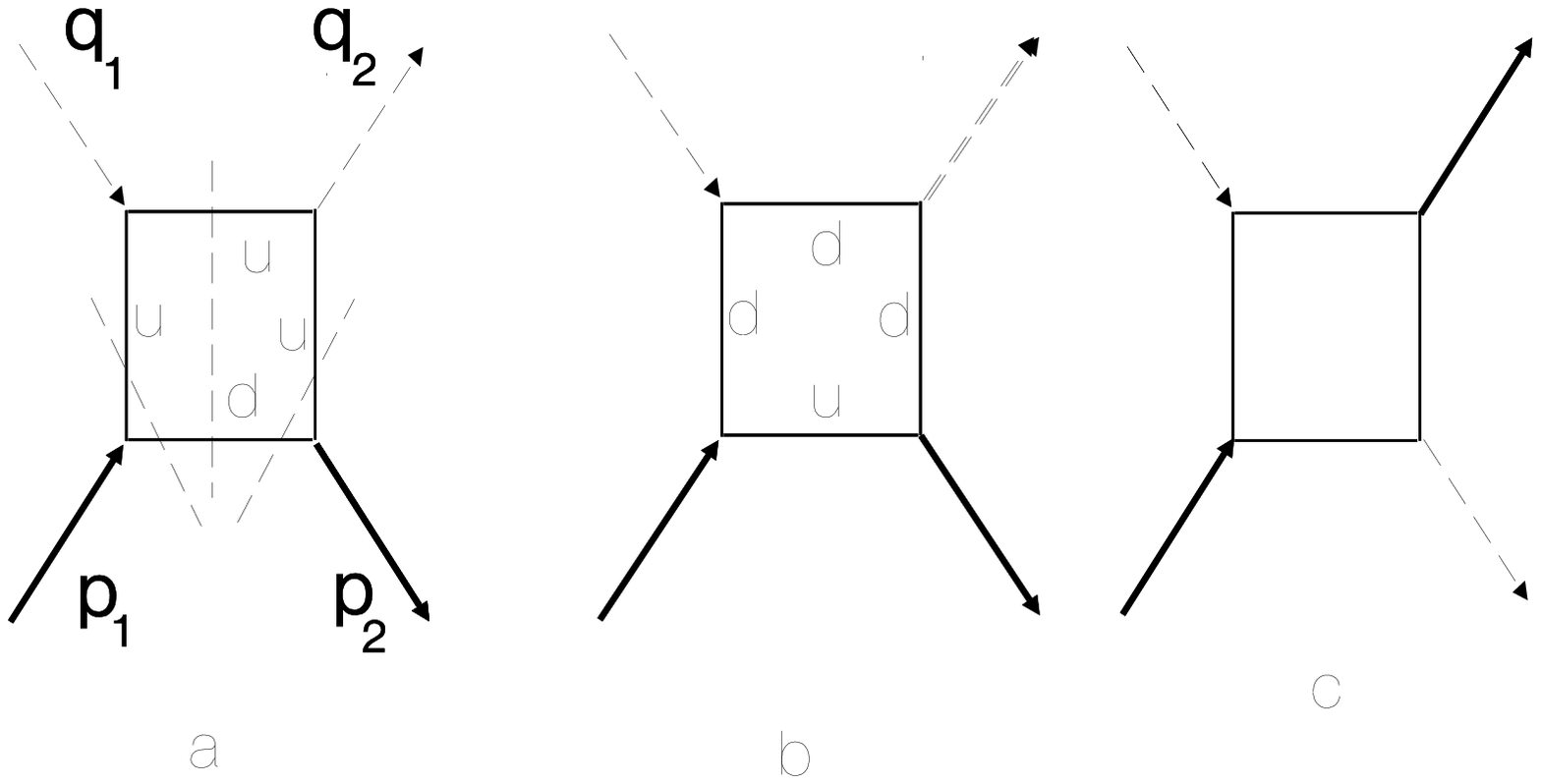} \caption{Diagrams,
corresponding to the unit operator
contribution. Dashed lines  with arrows correspond to the photon,
thick solid - to pion or rho current.}
\end{figure}

\begin{figure}
\epsfxsize=4cm \epsfbox{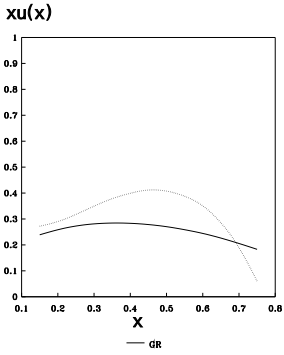} \caption{
Quark distribution function in pion (dashed line) and the fit
from [4b] - solid line.}
\end{figure}

\begin{figure}
\epsfxsize=7cm
\epsfbox{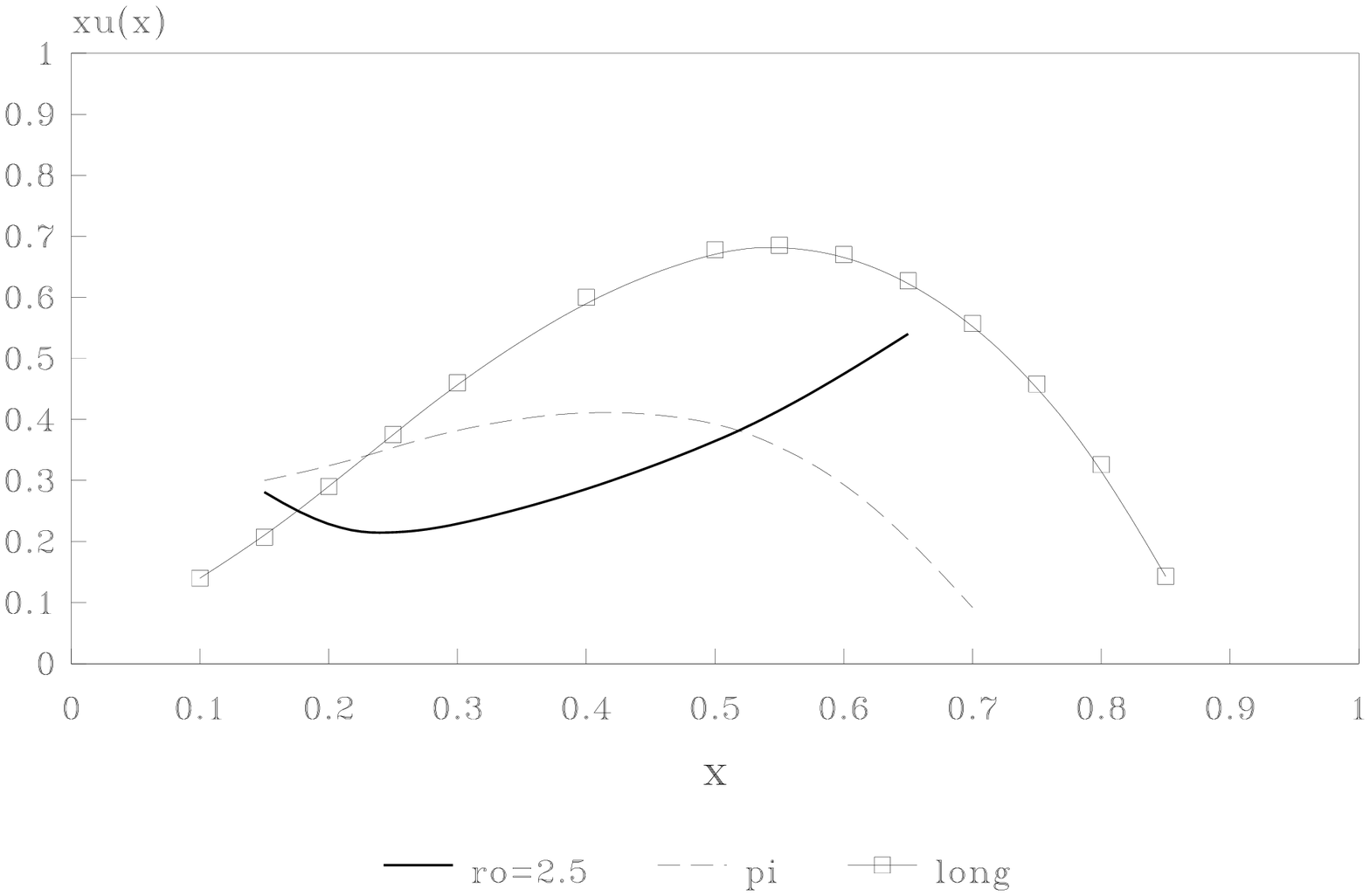}
\caption{Quark distributions in longitudinal $\rho$, (solid line with
squares), transversal $\rho$ (solid line) and pion (dashed
line)}
\end{figure}

\begin{figure}
\epsfxsize=4cm \epsfbox{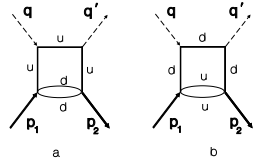} \caption{Bare loop
diagram in case of proton.}
\end{figure}

\begin{figure}
\epsfxsize=7cm \epsfbox{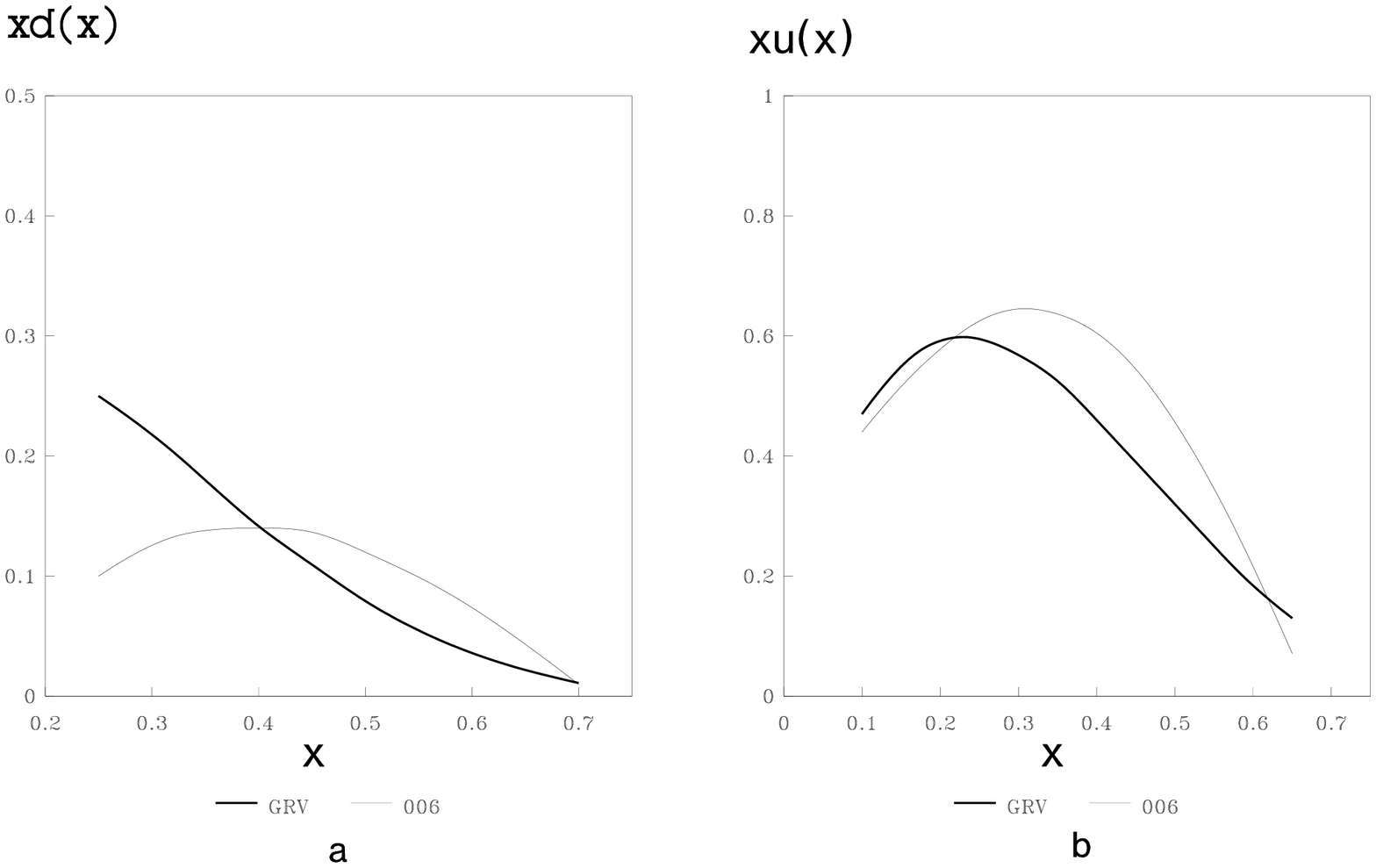} \caption{Valence $u$- and
$d$-quark distributions in proton (dashed lines) in comparison
with the fit [4a] of deep inelastic scattering data (solid lines)}
\end{figure}

\end{document}